\documentclass[10pt,conference]{IEEEtran}
\usepackage[left=0.67in,right=0.67in,top=0.75in,bottom=1.05in]{geometry}
\IEEEoverridecommandlockouts
\usepackage{cite, balance}
\usepackage{amsmath,amssymb,amsfonts}
\usepackage{enumitem}
\usepackage{graphicx}
\usepackage{textcomp}
\usepackage{subcaption}
\usepackage{dblfloatfix}
\usepackage[utf8]{inputenc}
\usepackage{url}
\usepackage{xcolor}
\usepackage{color}
\usepackage{multirow}
\usepackage{url}
\usepackage{algpseudocode}

\usepackage[T1]{fontenc}
\usepackage{newtxtext,newtxmath}

\usepackage[linesnumbered,ruled,vlined]{algorithm2e}
\def\BibTeX{{\rm B\kern-.05em{\sc i\kern-.025em b}\kern-.08em
    T\kern-.1667em\lower.7ex\hbox{E}\kern-.125emX}}

\begin{document}
\title{Energy-Efficient Satellite IoT Optical Downlinks \\ Using Weather-Adaptive Reinforcement Learning}

\author{
\IEEEauthorblockN{
Ethan Fettes\IEEEauthorrefmark{1},
Pablo G. Madoery\IEEEauthorrefmark{1},
Halim Yanikomeroglu\IEEEauthorrefmark{1},
Gunes Karabulut Kurt\IEEEauthorrefmark{1}\IEEEauthorrefmark{6},}
Abhishek Naik\IEEEauthorrefmark{2},
Colin Bellinger\IEEEauthorrefmark{2},
Stéphane Martel\IEEEauthorrefmark{4},
Khaled Ahmed\IEEEauthorrefmark{4},
Sameera Siddiqui\IEEEauthorrefmark{3}\\

\vspace*{0.3cm}

\IEEEauthorblockA{\IEEEauthorrefmark{1} Non-Terrestrial Networks (NTN) Lab, Department of Systems and Computer Engineering, Carleton University, Canada}
\IEEEauthorblockA{\IEEEauthorrefmark{6} Poly-Grames Research Center, Department of Electrical Engineering, Polytechnique Montréal, Montréal, Canada}
\IEEEauthorblockA{\IEEEauthorrefmark{2} National Research Council Canada, Canada}
\IEEEauthorblockA{\IEEEauthorrefmark{3} Defence Research and Development Canada, Canada}
\IEEEauthorblockA{\IEEEauthorrefmark{4} Satellite Systems, MDA, Canada}

}

\maketitle
\IEEEpubidadjcol

\begin{abstract}

 Internet of Things (IoT) devices have become increasingly ubiquitous with applications not only in urban areas but remote areas as well. These devices support industries such as agriculture, forestry, and resource extraction. Due to the device location being in remote areas, satellites are frequently used to collect and deliver IoT device data to customers. As these devices become increasingly advanced and numerous, the amount of data produced has rapidly increased potentially straining the ability for radio frequency (RF) downlink capacity. Free space optical communications with their wide available bandwidths and high data rates are a potential solution, but these communication systems are highly vulnerable to weather-related disruptions. This results in certain communication opportunities being inefficient in terms of the amount of data received versus the power expended. In this paper, we propose a deep reinforcement learning (DRL) method using Deep Q-Networks that takes advantage of weather condition forecasts to improve energy efficiency while delivering the same number of packets as schemes that don't factor weather into routing decisions. We compare this method with simple approaches that utilize simple cloud cover thresholds to improve energy efficiency. In testing the DRL approach provides improved median energy efficiency without a significant reduction in median delivery ratio. Simple cloud cover thresholds were also found to be effective but the thresholds with the highest energy efficiency had reduced median delivery ratio values.
\end{abstract}


\section{Introduction}
The Internet of Things (IoT) is a rapidly growing area of the space economy. IoT devices located in remote areas such as forestry and agriculture are frequently served via satellite due to the high cost of installing terrestrial network infrastructure. Two major trends have occurred in recent years: the quality and quantity of IoT devices deployed to remote areas have rapidly increased and low Earth orbit (LEO) satellites have become increasingly popular for servicing IoT devices due to lower cost \cite{SIoT_Survey}.\\

As the quality of sensors serviced by these satellites has increased, a major challenge is transmitting all the collected data to the ground via the spacecraft communication payload. In recent years free space optical (FSO) communications have been proposed due to the high data rates that can be achieved. However, FSO communications are heavily impacted by adverse weather conditions such as cloud cover. As cloud cover can be forecasted using physical weather models such as the regional deterministic prediction system (RDPS) \cite{RDPS} or machine learning-based techniques \cite{cloud_prediction}, using this information to make intelligent routing decisions is possible. \\

During communication windows between satellites and ground stations via FSO links, both coarse pointing, acquisition, and other supporting processes must take place to allow for successful communication \cite{CubeSat_FSO_Terminal}. All of these services require energy provided by the spacecraft power system and unnecessary expenditures are undesirable due to the power-limited nature of spacecraft \cite{Power_Budget_Operations}.\\

Hence, developing strategies to use more favourable transmission periods (frequently referred to as contacts) when possible is advantageous. In this work, the benefits of using cloud cover forecast data to improve energy efficiency while maintaining the same delivery ratio of customer data are analyzed. Both an advanced scheme using deep reinforcement learning (DRL) and simple schemes using basic cloud cover thresholds are evaluated. Specifically, the impact of initial data volume is evaluated and a case study is performed using historical weather data to determine performance in realistic conditions. Subsequently, future work is proposed regarding potential  algorithms and incorporating increased realism into the system model. These efforts resulted in the following contributions:\\

\begin{itemize}
    \item We propose a DRL approach using multiple deep Q-networks (DQN) agents for improving energy efficiency in the presence of weather disruptions for non-geostationary orbit (NGSO) satellites.
    \item We evaluate the proposed algorithm against basic heuristic and baseline routing schemes. We also identify other potential schemes that could be investigated in future work.
    \item We identify potential future improvements to the system model, particularly with respect to weather and energy modelling. 
\end{itemize}

The rest of the paper is structured as follows. Section \ref{sec:background} provides an overview of previous work in this domain, followed by Section \ref{sec:systemmodel} which outlines the system model, the proposed schemes, and the performance objectives. Section \ref{sec:evaluation} contains an analysis regarding the effect of data volume and a realistic case study that uses historical weather data to evaluate the proposed schemes. Lastly, Sections \ref{conclusion} and \ref{futurework} discuss important takeaways from the previous sections and outline potential avenues for future research.

\section{Background and Related Work}
\label{sec:background}

LEO satellites have become increasingly important in the provision of satellite IoT (SIoT) services \cite{SIoT_Survey}. In addition due to cost constraints sparse constellations SIoT network configurations generally consist of single satellites or sparse constellations \cite{Sparse_Constellations} \cite{SIoT_Survey}. Frequently, SIoT satellites do not have inter-satellite connectivity resulting in only intermittent connectivity with the ground segment.\\

To establish FSO communications between satellites and the ground, a series of processes termed acquisition, pointing, and tracking \cite{CubeSat_FSO_Terminal} must be conducted. In satellites, these processes require a number of sensors, actuators, and other hardware to be activated. This results in increased power consumption that must be provided by power from photovoltaic cells, secondary batteries, or both,\cite{Spacecraft_Battery_Paper}. Battery energy storage is limited on satellites due to space and mass constraints \cite{Battery_Review_Article}.\\ 

One of the major differences between FSO-based satellite-to-ground communications and RF-based communications is the higher vulnerability to adverse weather conditions. Space-to-ground FSO links are affected by two main effects, mie scattering and geometric scattering \cite{FSO_link_budget}. Mie scattering is due to cloud cover, while geometrical scattering is associated with fog, dust, and other low-altitude weather conditions \cite{FSO_link_budget}. Frequently, attenuation due to cloud cover vastly exceeds the link margin with attenuation values greater than 100 dB per km possible \cite{Ground_Station_Optimization}. Hence, the presence of cloud cover in the line of sight (LoS) between the satellite and the ground station can be assumed to cause link blockages \cite{Ground_Station_Optimization}.\\

The use of weather forecast information for satellite operations and communications have been proposed in several existing works. In \cite{TwostageEOschedule} the authors address the issue of scheduling Earth observation (EO) collections, and the subsequent downlink of collected information in the presence of weather conditions that can impede EO. While in \cite{LinkBudgetWeatherPred}, weather predictions are used to improve the efficiency of RF downlinks by more accurately identifying favourable contacts. However, this work does not take into account quality of service metrics as it focuses solely on optimizing link parameters. In terms of FSO communications the most common method for mitigating the challenge of adverse weather conditions is to make use of site diversity as proposed in \cite{Site_Diversity_Clouds}. However, this approach fails to address the issue of utilizing contacts that are inefficient. Additionally, adverse weather can be mitigated by adding new network architectures but at high capital cost \cite{ethanspaper}. This gap is addressed by schemes presented in this work from the perspective of energy efficiency.
\section{System Model and Problem Formulation}
\label{sec:systemmodel}

\subsection{System Architecture}
\begin{figure}
\centering
\includegraphics[width=0.98\linewidth]{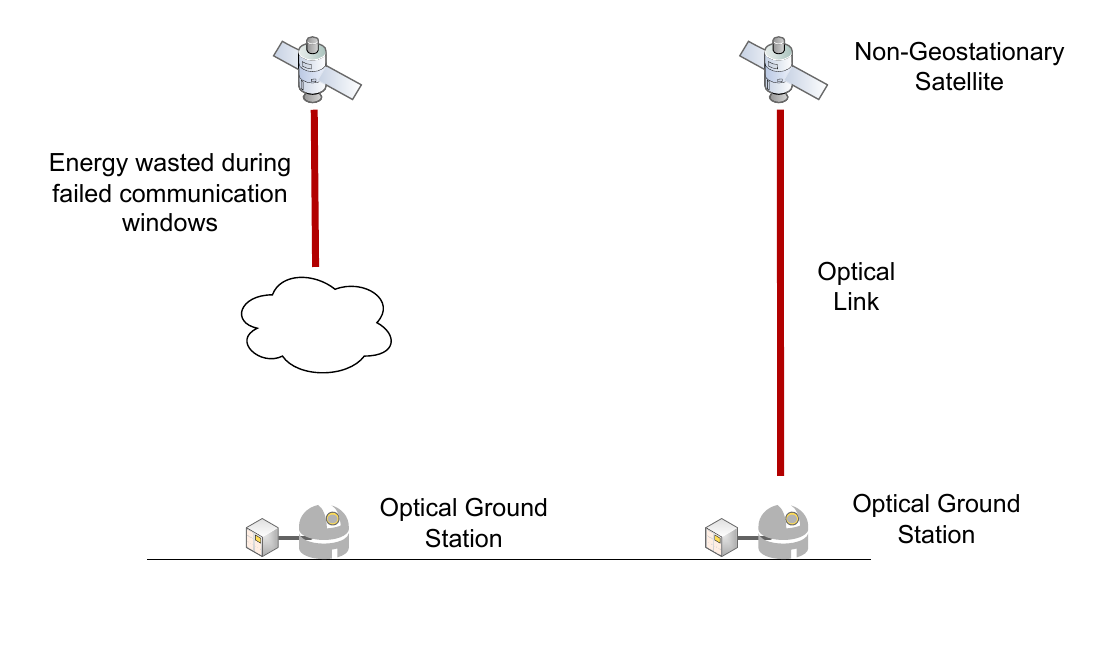} 
  \caption{System architecture.}
  \label{fig:Sys_Arch}
\end{figure}

The system model shown in Figure \ref{fig:Sys_Arch}, outlines the key aspects of the problem associated with efficient optical downlink in the presence of weather conditions. We assume only a single satellite is used and communication with the remote IoT devices is conducted before the simulation. It is also assumed that only one ground station is available at any one time. Given this architecture, the data must be stored and forwarded by the satellite necessitating the use of delay tolerant networking (DTN) protocols. The most popular DTN protocol for satellite networks is bundle protocol which packages data into bundles \cite{CGRTutorial}.\\
As we assume a NGSO satellite is used, the total system capacity without weather impairments can be defined as
\begin{equation}
    V=\sum_{i=1}^{n} (\text{Contact Volume}_i).
\end{equation}

To model the impact of adverse weather conditions, the space-to-ground communication channel is modelled as an on-off channel. The status of the channel, and hence availability of a ground station during a contact is dependent on the position of the satellite relative to the clouds and the ground station. This position changes rapidly due to the satellite orbit. Availability for each contact is defined as $\Delta$ with units defined in terms of data volume (the amount of data that can be delivered during the specific contact).\\

As spacecraft power usage modelling is outside the scope of this work, a simple model is used to measure excess power consumption. The power model is limited to communications power rather than total spacecraft power. We define excess power consumption as the number of unsuccessful bundle transmissions due to cloud cover. Given excess power consumption $a$, contact length $b$ (measured in bundles), and number of delivered bundles $d$, 
\begin{equation}
a=b-d.
\end{equation}

This assumes that the acquisition and pointing processes required to re-acquire communication between the satellite and the ground station are constant during the link failure periods. Excess power consumption can be eliminated from a contact with link failure by disabling the laser terminal pointing and acquisition functions but this also eliminates the ability to transmit if conditions improve during the contact. Finally, it is assumed that there is an acknowledgement mechanism present that notifies transmitting nodes if bundles are received.
\subsection{Problem Formulation}
\paragraph*{Objectives}
Given the system model described above, the two key metrics of interest are delivery ratio and energy efficiency. Delivery ratio is defined as the number of bundles delivered to the ground station, divided by the number of bundles that are scheduled to be sent during the transmission window. Energy efficiency is defined as the number of bundles delivered, divided by the contact length in bundles. The algorithms below assume that delivery ratio is the most important metric for satellite operators as delivering the maximum amount of sensor data is generally a key objective. The secondary objective is to improve energy efficiency. These objectives can be represented as follows:\\
\begin{equation}
    z=\alpha \times w+(1-\alpha)\times y.
\end{equation}
Where $\alpha$ is a weight value, $w$ represents the delivery ratio and $y$ is the energy efficiency.
\paragraph*{Threshold Schemes}
A simple way to reduce excess energy usage during transmission periods is by applying thresholds based on the cloud cover forecast. Given a cloud cover threshold value of $\nu$ and a current cloud cover percentage $\lambda$ the scheme logic can be described as follows:
\\
\begin{equation}
\text{Contact Decision}= 
\begin{cases} 
\text{Use Contact}  & \text{if } \lambda <=  \nu\\
\text{Remove Contact} & \text{if } \lambda > \nu\\
\end{cases}.
\end{equation}
The effect of different cloud cover thresholds depends on the distribution of weather conditions. The obvious benefit of threshold schemes is their low computational cost and in many cases, calculations can be done on the ground. This can be particularly beneficial for resource-constrained systems such as CubeSats. 
\paragraph*{DRL Agent Definition}
The sequential selection of contacts fits well with Markov decision processes and hence, DRL is an attractive approach. As ground stations are assumed to be far enough apart to not allow simultaneous communications, the action space is restricted to:
\begin{equation}
\mathbf{Actions} \in \{0, 1\}.
\end{equation}

It is assumed that the DRL agent is located on the satellite node to allow for access to data storage information. The observation information available to the agent for each decision is as follows:\\
\begin{equation}
\mathbf{Observation Space} = \begin{bmatrix}
\text{$\lambda \quad$} 
\text{$\Omega \quad$} 
\text{$\theta \quad$} 
\text{$\Lambda$}
\end{bmatrix}^T,\\
\end{equation}
where $\lambda$ is the cloud cover for the next contact, $\Omega$ is the remaining data volume, $\Omega$ is the difference between the total bundles and the number of delivered bundles, $\theta$ is the remaining system capacity given $m$ is the current contact,
\begin{equation}
\theta=V-\sum_{i=1}^{m} (\text{Contact Volume}_i),
\end{equation}
and $\Lambda$ is a matrix with all future weather and contact information for the downlink period. DQN is chosen as the reinforcement learning algorithm due to its ability to handle larger state-action spaces than Q-Learning\cite{dqn}.

The objectives of the DRL agent are to maximize the delivery ratio and where possible improve energy efficiency. To ensure that fulfilling these objectives is desirable for the agent, the reward function is defined as follows. At each step, the agent chooses an action from the action space. If the agent chooses to not utilize a contact, the agent receives zero reward. Otherwise, the agent receives a reward defined by the number of bundles delivered during the current time step and how efficiently these bundles are delivered. It is ensured that in the step reward, if the agent successfully delivers bundles, the reward will be positive. The only circumstance that results in a negative reward for  the agent is if a contact is selected and no bundles are delivered. Improving energy efficiency is rewarded by allocating increased positive rewards for using contacts that are more efficient. The episode component of the reward function rewards the overall delivery ratio and energy efficiency. These components are outlined below:
\begin{equation}
\text{Episode Return}=\text{Episode Reward}+\sum(\text{Step Reward}),
\end{equation}
The delivery ratio and total utilized contact time for an episode are  defined as $\eta$ and $\Theta$. And with the scaling factor $c$, they define the episode reward,
\begin{equation}
\text{Episode Reward}= 
\begin{cases} 
  2 \times c \times \eta \times \Theta& \text{if } \eta ==1,\\
c \times \eta & \text{if } \eta <1. \\
\end{cases}
\end{equation}
The step reward consists of components $F1$,$F2$,$F3$:
\begin{equation}
F1=\frac{c}{\beta}\times\omega,
\end{equation}
\begin{equation}
F2=\frac{c}{\beta}\times\omega\times\frac{\epsilon}{\epsilon+\zeta},
\end{equation}
\begin{equation}
    F3=-\epsilon \times \frac{c}{2 \times\beta},
\end{equation}
\begin{equation}
\text{Step Reward}= 
\begin{cases} 
F1-F2  & \text{if } F1 > 0 \text{ and Action=1}, \\
-F3 & \text{if } F1 = 0 \text{ and Action=1},\\
0 & \text{if } \text{Action}=0.\\
\end{cases}
\end{equation}
In the equation above, $\beta$ is the data volume that needs to be delivered during the transmission window, $\omega$ is the number of bundles delivered during the current time step, $\epsilon$ is the number of unsuccessful transmissions during the current time step, and $\zeta$ is the length of the contact. \\

This DQN configuration is implemented in MATLAB using the reinforcement learning toolbox and the agent hyperparameters are set as described in Table \ref{Hyper_table}. During experimentation, it was found that a single DQN agent suffered from poor performance in certain data volume ranges. A solution to this issue is the use of multiple DQN agents each trained on a particular range of data volume values. During implementation and testing, the appropriate agent is selected based on the data volume value using a simple switch architecture.\\
\begin{table}
\centering
\caption{Hyperparameter tuning configuration.}
\begin{tabular}{|c|c|}
\hline
Hyperparameter & Value\\
\hline
Learning Rate & 0.01\\
\hline
Experience Buffer Size & 10000\\
\hline
Discount Factor & 0.99\\
\hline
Epsilon Decay & 0.0050\\
\hline
MiniBatch Size & 64\\
\hline
Target Update Frequency & 1\\
\hline
L2 Regularization & 0.0001\\
\hline
Target Smooth Factor & 0.001\\
\hline
Double DQN & True\\
\hline
\end{tabular}
\label{Hyper_table}
\end{table}

\section{Performance Evaluation}
\label{sec:evaluation}
\subsection{Total Data Volume Impact Analysis}
\begin{figure*}[htbp]
    \centering
    \begin{minipage}[b]{0.35\textwidth}
        \centering
        \includegraphics[width=\linewidth]{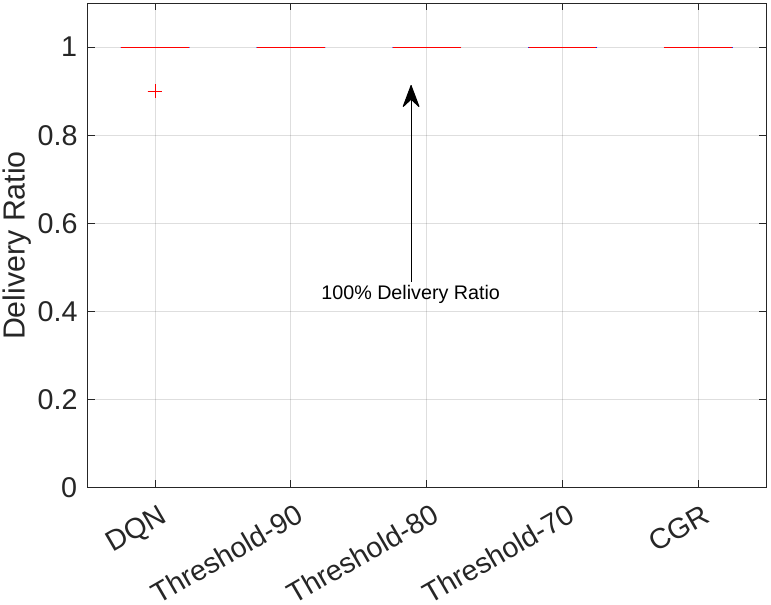}
        \subcaption{Delivery ratio- data volume 10 \% of system capacity.}
        \label{fig:packet1}
    \end{minipage}
    \hspace{0.5cm} 
    \begin{minipage}[b]{0.35\textwidth}
        \centering
        \includegraphics[width=\linewidth]{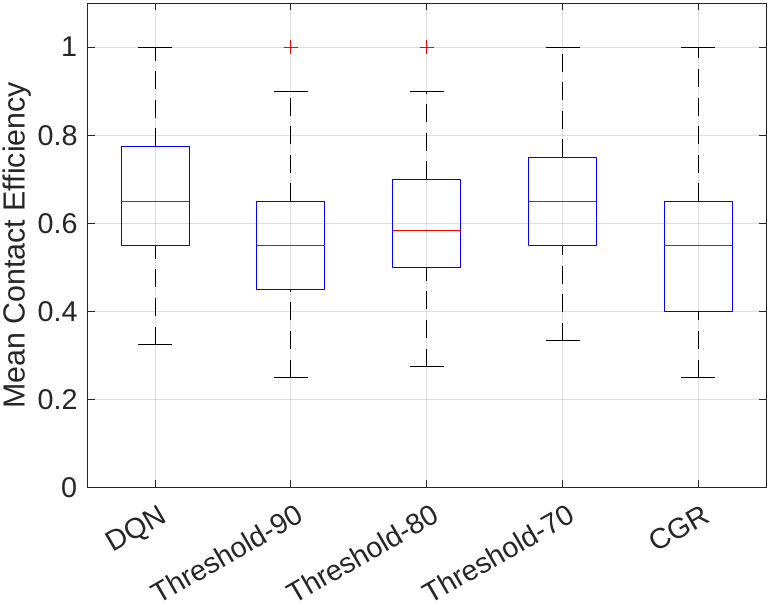}
        \subcaption{Mean contact efficiency- data volume 10 \% of system capacity.}
        \label{fig:packet2}
    \end{minipage}

    \vskip\baselineskip 

    \begin{minipage}[b]{0.35\textwidth}
        \centering
        \includegraphics[width=\linewidth]{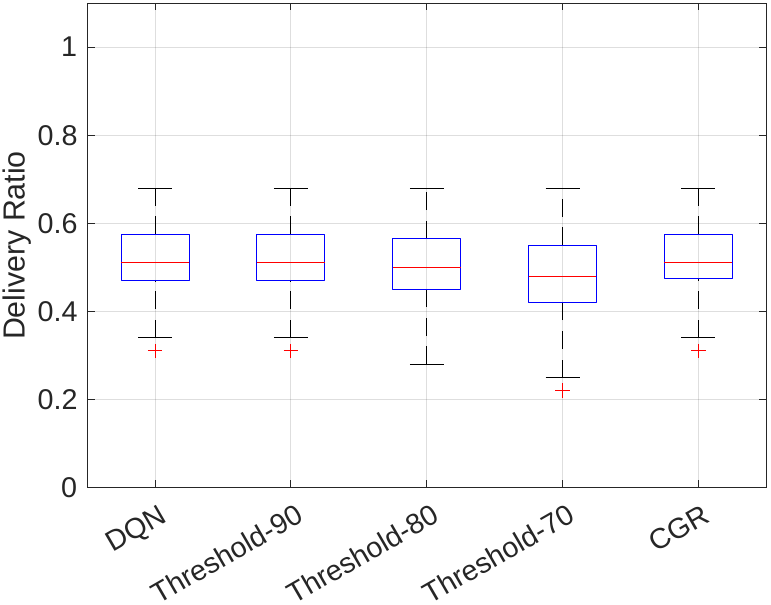}
        \subcaption{Delivery ratio- data volume 100 \% of system capacity.}
        \label{fig:packet3}
    \end{minipage}
    \hspace{0.5cm} 
    \begin{minipage}[b]{0.35\textwidth}
        \centering
        \includegraphics[width=\linewidth]{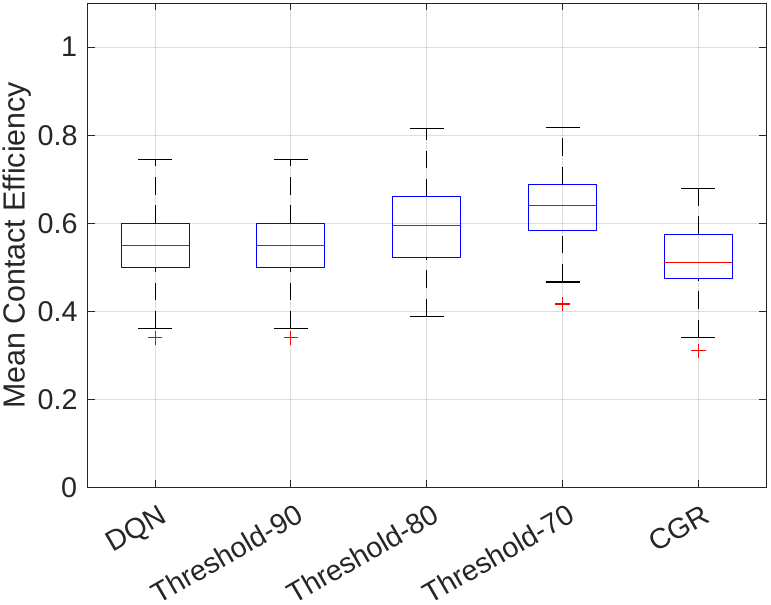}
        \subcaption{Mean contact efficiency- data volume 100\% of system capacity.}
        \label{fig:packet4}
    \end{minipage}
    \caption{Data volume volume testing was conducted to determine the impact of varying the initial data volume distribution has on the performance of the proposed DQN and threshold schemes.}
    \label{fig:Packet_tests}
\end{figure*}

An important factor in the optimal strategy for the DRL scheme and the optimal threshold value is the ratio between the initial value of $\Omega$ and the available capacity $V$ of the system. The proposed schemes are tested with two different $\Omega$ values and a set of 10 equal length contacts. 
\begin{equation}
    V=\sum_{i=1}^{10} \text{Contact Volume}.
\end{equation}
In terms of the weather and the impacts on $\Delta$, a cloud cover value $\mu$ is defined as,
\begin{equation}
\mu= \text{Fractional Cloud Cover} \times \text{Contact Length},
\end{equation}
and is used to define $\Delta$ as normally distributed around the expected number of bundles delivered:
\begin{equation}
\Delta \sim N(\mu, \sigma^2).
\end{equation}
As evaluating the impact of data volume is the primary objective for this test cloud cover values are uniformly distributed and rounded to the nearest multiple of 10 for simplicity. Due to all contacts being equal length, the value $\Lambda$ is simplified to a vector containing only future weather information.\\
In Figure \ref{fig:Packet_tests}, the performance of different threshold values, the DRL scheme and a baseline DTN routing scheme, contact graph routing (CGR) \cite{CGRTutorial} are shown for different data volume values. The DRL scheme performs almost identically to the 90 percent threshold scheme when data volume is equal to the system capacity. For low data volume values the DRL scheme is able to outperform both the 80 percent and 90 percent threshold schemes in terms of energy efficiency while maintaining delivery ratio.
\subsection{Case Study}
To determine the performance of our system in more realistic circumstances we define a realistic configuration where a single satellite transmits data to ground stations located in Ottawa and Calgary, Canada. To improve realism several modifications are made to the system model. The lengths of contacts are determined assuming a single satellite at 500 km altitude in polar orbit over random 3 day periods in 2024 assuming the selected ground station configuration. While the data volume values are uniformly distributed between five and 100 percent of the $V$, historical weather conditions are obtained from \cite{meteoblue}. The impact of cloud cover on ground station availability is modelled by taking samples every 20 seconds to determine if the ground station is obstructed by clouds. To simulate this sampling process we define a random variable $Z$ as follows:
\begin{equation}
    Z \sim U(0, 1).
\end{equation}
$Z$ simulates a sample for LoS access. Hence, given a cloud cover $p$ we can determine the status of the link:
\begin{equation}
    p=\text{Fractional Cloud Cover},
\end{equation}
\begin{equation}
    \text{Ground Station Available}=
    \begin{cases}
        1 & \text{if } Z>p,\\
        0 & \text{if } Z \leq p, \\
    \end{cases}
\end{equation}
where $\Delta$ is the sum of these samples for a given contact.\\ 
The DRL configuration remains the same except that the observation info $\Lambda$ is expanded to include the contact length info. The DRL configuration is trained on a separate ground station configuration of Calgary and Inuvik, Canada. In addition to the schemes tested in the previous experiment, a multi-threshold scheme that uses different thresholds for specific data volume ranges is also tested. This solution attempts to mitigate the impact of different data volume values. The thresholds used in this scheme are selected using 2023 weather data from \cite{meteoblue} for Ottawa and Calgary. For the purposes of this case study, bundles assumed to be 20 GB in size while the data rate of the link was assumed to be 8 Gbit/s.\\

\begin{figure*}[htbp]
    \centering
    \begin{minipage}[b]{0.35\textwidth}
        \centering
        \includegraphics[width=\linewidth]{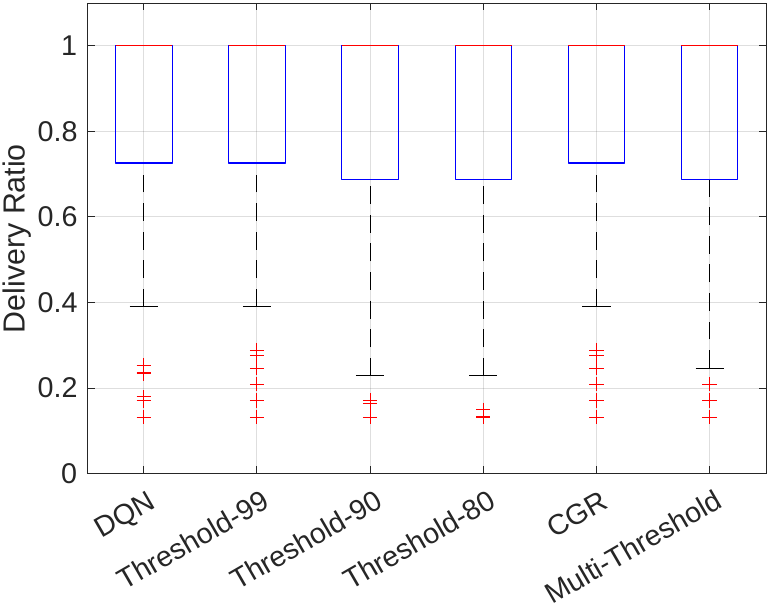}
        \subcaption{Delivery ratio for Ottawa-Calgary configuration.}
        \label{fig:sens_1}
    \end{minipage}
    \hspace{0.5cm} 
    \begin{minipage}[b]{0.35\textwidth}
        \centering
        \includegraphics[width=\linewidth]{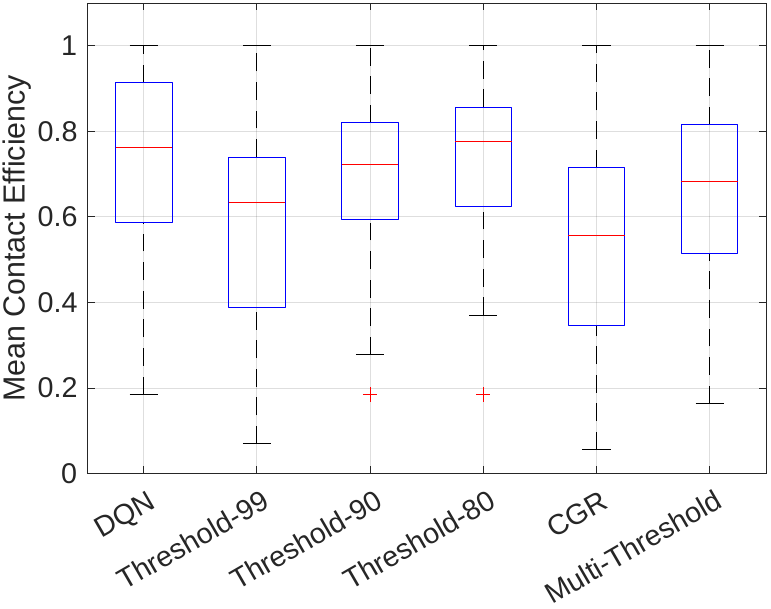}
        \subcaption{Mean contact efficiency for Ottawa-Calgary configuration.}
        \label{fig:sens_2}
    \end{minipage}

    \vskip\baselineskip 
    \caption{Results for Ottawa-Calgary ground station configuration with single LEO satellite.}
    \label{fig:Sensitivity_Testing}
\end{figure*}

As shown in Figure \ref{fig:Sensitivity_Testing}, the DRL scheme shows improved performance over all threshold schemes. This validates that the DRL scheme can perform well in more realistic configurations. Additionally, as the DRL scheme is trained on a different ground station configuration than the Ottawa-Calgary configuration, these results also show that the DRL scheme can perform well on ground station configurations that it has not been trained for. This is an important distinction to threshold schemes where each ground station configuration needs to be allocated a specific threshold to meet performance objectives.
\subsection{Discussion}
The results of both the data volume analysis and the case study show improvements in performance using the DRL scheme versus threshold schemes. Although lower thresholds outperform the DRL scheme in terms of overall energy efficiency, they sacrifice some delivery ratio performance to achieve this. Hence, DRL is most suitable for capable satellites with sufficient computational capacity to conduct inference operations onboard, that have strict performance requirements. For satellites with looser efficiency and delivery ratio requirements, thresholds are likely more appropriate due to their low computational complexity.\\

Another use case for DRL is for satellite systems that require flexibility, particularly in the ground station sites that are used. This is due to the high impact that site cloud cover distributions have on all but the most conservative threshold scheme. If ground station sites are changed or new ground station sites are added during a mission, threshold scheme would likely need to be manually tuned as they may no longer result in performance requirements being met. 
\section{Conclusion}
\label{conclusion}
The results obtained in previous sections demonstrate the DRL scheme provides improved energy efficiency when compared to basic threshold schemes that achieve the same delivery ratio. It was also shown that the DRL scheme is able to meet performance objectives when different initial data volume values are used. Finally, it was shown that the DRL scheme does not need to be trained for a specific ground station configuration, allowing for increased ground segment flexibility.
\section{Future Work}
\label{futurework}

In terms of future work, there are several notable improvement that can be made to both the modelling and experiment design. Firstly, as the energy model used is very simple, improvements to this model would increase the applicability of this work. Secondly, incorporating uncertainty in the weather forecast information would be an important step in determining if the DRL scheme is able to operate effectively in the real world. Thirdly, as only DRL was evaluated for this work, a comparison with other schemes such as Monte Carlo or dynamic programming would be beneficial.
\section*{Acknowledgements}
This work has been supported by the National Research Council Canada's (NRC) High Throughput Secure Networks program within the Optical Satellite Communications Consortium Canada (OSC) framework, MDA and Mitacs. Additionally, we would like to thank meteoblue for providing the historical weather data used in this work.

\end{document}